# Smartbanner: Intelligent banner design framework that strikes a balance between creative freedom and design rules

Guandong Li[1*], Xian Yang[1]

1.*Suning, Xuanwu, Nanjing, 210042, Jiangsu, China.

*Corresponding author(s). E-mail(s): leeguandong@gmail.com

*Abstract*: Companies use banners extensively to promote their products, and the intelligent automatic synthesis of banners is a challenging event. Under the premise of inputting only a small amount of information such as product, text and size, it can synthesize styles with high freedom and richness, but at the same time, it must satisfy the design specifications of advertisers for advertising and scenes. We propose an intelligent banner design framework that strikes a balance between creative freedom and design rules, called smartbanner. Smartbanner consists of planner, actuator, adjuster and generator. The banner is synthesized through the combined framework, which fully liberates the designer and reduces the threshold and cost of design. It increases the click-through rate by 30%, improves the human efficiency of designers by 500% under the condition of ensuring the quality of creation, and synthesizes hundreds of millions of pictures in batches throughout the year.

1.Introduction

Banner design is ubiquitous in modern life. Creating and refining designs can be time-consuming and require expertise to clearly convey information while satisfying aesthetic needs. Although original designs are more artistic, there are some measurable rules that must be followed to create a series of designs with a consistent visual appearance. When advertising, different advertisers often put forward their own advertising requirements for the design of banners, and they are placed in different locations or scenarios, such as information flow, joint banner and other advertising spaces, header images, social sharing applets, etc. Different scenarios and placements require designers to repeatedly modify the materials of the same theme. Even banners of the same style will take a lot of time, and different advertising positions will also constrain the design. Of course, for designers, when a creative new banner needs to be completed, it may be difficult and time-consuming to create an effective and design-compliant diagram [1,2]. Given product images and textual descriptions, designers need to consider a number of design elements to meet aesthetic requirements. With the growing demand for creative design, intelligent design can not only greatly reduce the workload of designers, but also generate new ideas and inspiration for designers.

Banner is usually composed of background, product, text, ornament, logo, text binding mask and product binding mask,etc. The text includes the main title, subtitle and action words, and the product may also be one or more. Banner design is a complex problem, and it is difficult to have an end-to-end design solution that is both creative and in line with design specifications. There was a topological transformation based on style before [3]. The consequence of this style transformation style is that the layout and style of the design look relatively simple. Compared with the original image, it may only be a change in the layout of the left and right or up and down, which does not reflect the creative ability. There are also some techniques for automatically generating graphic designs, but they mainly focus on arranging user-specified layouts of images and text [4, 5, 6]. In the design of creative banners, two issues must first be considered. The first is to solve the layout problem, how to generate a template that matches the product aesthetically and semantically is very important. The second is the layout and fine-tuning of the core modules, including the layout of the text area and the product area and the adjustment of the matching element mask. In the e-commerce scenario, text is the core slogan. Many touch points of products to the crowd are hidden in the text, and the layout of the text is particularly important. Multiple product arrangement is also one of the keys to affecting banners. Their element layout, the combination of bound ornaments and the selection of mask should reflect that the selected context is consistent with the semantics of product image, and ensure that the color, shape and size of design elements are coordinated with each other. In addition to this, two other factors hidden in the banner design are also crucial. First, banner's creativity comes not only from the combination of design elements and templates themselves, but also from the historical click-through rate. Therefore, it is considered to integrate the click through rate information into the creative design, and to fine-tune and optimize the generated banner according to the rolling of scene data [7,8]. Secondly, creative banners often only enter the product text and size, and can automatically generate creative maps. It is a process with a high degree of freedom in design. There can be many results in the layout of layout elements. This process requires high diversity. However, advertisers need banners with regular output. Our algorithm strikes a balance between degrees of freedom and rules, and the output graph is widely used in information flow and the suning.com.

Our banner creative design only needs to input the product text and size, and can automatically generate a creative banner, which greatly liberates the design productivity, we call it smartbanner. Smartbanner consists of planner, actuator, adjuster and generator. The planner selects templates in the template library through template recall, template pre-ranking and creative ranking to find a good layout. Creative ranking is a supervision model integrating information such as image , text and historical click through rate. A good layout is half the success. The actuator consists of a size extension tool, a text layout model, and multi-product template rules. The adjuster maintains an element retriever and a set of fine-tuning algorithms to adjust and replace elements in the template. The generator combines images based on the finely adjusted banner. Smartbanner is an intelligent design framework that takes into account the degree of freedom and design rules, which not only ensures good creativity, but also ensures the normative design. In the e-commerce scenario, the normative design is very important. Smartbanner is also a combined design framework that supports almost all automated intelligent design services of Suning.com. In advertising, designers only need to design a small number of basic templates to generate massive banners. Through better product creative collocation, the click-through rate can be increased by 30%. Under the condition of ensuring the quality of creation, the efficiency of designers will be improved by 500%, and hundreds of millions of pictures will be synthesized in batches throughout the year.

Our contributions in this paper are:
1. The smartbanner is proposed. Smartbanner is an intelligent design framework that takes into account the degree of freedom and design rules. You only need to input product text and size information to complete the banner design. Consists of planning, actuator, adjuster and generator. It not only ensures good creativity, but also ensures the normative design, liberating designers.
2. The click rate information is integrated into the creative design, and the generated banners are arranged and optimized according to the scrolling of the scene data.
3. A text layout model designed by a neural network is introduced into the actuator, which supports multi-product input and size expansion modules, does not forcefully rely on the recalled template size, and can support banner output of any size.

2.Related works

Tang YC[9] emphasized the core role of the domain knowledge database in the theoretical model of design intelligence. The domain knowledge database represents the data of design, style, art, technology, etc., and is the data source of the creative module. Microsoft[6] proposed a computing framework with a set of theme related layout templates and an integration of high-level aesthetic design principles (in a top-down manner) and low-level content features (in a bottom-up manner) . The layout template is designed by using the prior knowledge of domain experts to define spatial layout, semantic color, harmonic color model, font emotion and size constraints. This framework formulates typography as an energy optimization problem by minimizing the cost of text intrusion, waste of free visual space, and mismatch of information importance in perception and semantics. This work is quite instructive and forward-looking, and we learned a lot from it, including ideas from template layouts.

Much research has focused on developing optimal layout algorithms that can automatically place images and text in appropriate positions based on certain aesthetic criteria [10]. De signscape [11] is a representative example. It is an interactive system used to generate layout suggestions for a group of images and text input by users. To enhance the interactivity of the automatic layout generation process, Todi et al. proposed SketchPlore [12], a real-time layout optimization tool that automatically infers designer actions and uses predictive models to suggest more feasible layouts for generating user interfaces. Tabata et al. [13] proposed various candidates for automatically creating magazine layouts, preserving the original design style given user input of text, images, and page numbers. A recent work by Zheng et al. [14] is able to lay out images on text documents according to the content of paragraphs. The design of advertising posters usually starts with a very limited number of design elements specified by users. The algorithm needs to further select the design elements that match the product image, and correctly arrange all elements on the canvas to complete the design. To produce better stylized graphic designs, Jahanian [15] proposed a graphic design guide for magazine covers and designed a framework consisting of three main modules, including the layout of cover elements, color, and typography of cover lines. Yang et al. [16] designed a system to automatically generate digital magazine covers by summarizing a set of topic-related templates and introducing a computational framework that incorporates key elements of layout design. These works are usually rule-based and less intelligent. Vinci [17] proposed a design space to describe design elements in advertising posters, and introduced design sequences to feed back the design decisions of human designers when creating posters.

In the industrial world, Alibaba's Luban[18] and JD.com's Linglong system are extremely relevant to our work. Smartbanner focuses on large-scale implementation in actual scenarios, and the core is to strike a balance between creative freedom and design rules.

3.Methods

Smartbanner consists of planner, actuator, adjuster and generator. The planner is to get a good layout design, with the help of template recall,pre-ranking and creative ranking. After obtaining the layout design template, the actuator will layout the product and text areas in the banner respectively. The adjuster builds an evaluation algorithm based on aesthetic standards and some design rules, and further fine-tunes the elements, products, and text after layout. Finally, there is the generator, which renders the completed template into a graph. The overall flow chart is shown in Figure 1.

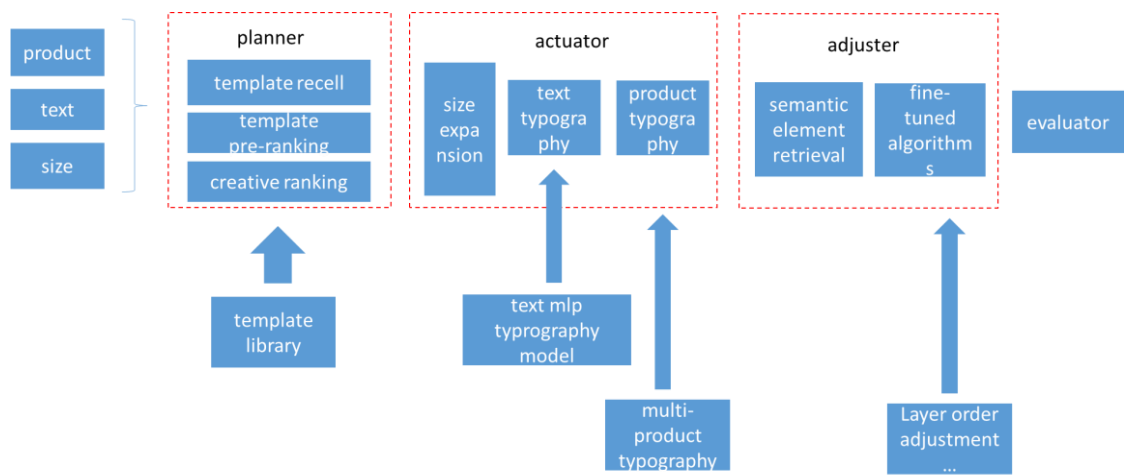

Fig 1. Smartbanner process diagram

3.1 planner

How to get the layout? A good layout is half the battle. We can directly generate layouts through the end-to-end method of layoutgan[4]/dcgan[19], but this way of generating layouts often faces two major drawbacks. First, the end-to-end generation is too free, the effect is unreadable, and the generated layout contains too many badcases, which cannot meet the design needs of advertisers. Second, the labeling requirements are high. In the case of limited resources, it is difficult to give a relatively good layout in the initial stage. However, the overfitting layout does not satisfy the diversity of the design. Therefore, under the balance of freedom and design rules, we chose the layout generation method of the template library. Predefined templates can provide relatively standardized design layouts. By adding design constraints to the template dimension, standardized elements and layout designs that meet the requirements of advertisers can be obtained. In addition, for different scenarios, we can provide corresponding relationships between different business units and template groups, limit the freedom of creativity and ensure the high availability of generated results.

3.1.1 template design

To introduce the difference between domain-specific layout design and computing content features, we add a narrower degree of freedom restriction to the template design to ensure the normative layout at the beginning of banner generation. The template designed by the designer uses more standardized elements:

a. In PSD, we limit the types of special effects and layers, and we classify the layer names in detail. These detailed classifications are conducive to the action and fine-tuning module to adopt different processing methods for layers with different types of names.

b. The template designed by the designer needs to add label information such as the type and style of the template, which helps us to train the label model to design the label of the feature dimension for the new template.

c. In order to ensure the richness of the template, we have made constraints on the number of products in the template, the number of text lines, the left and right layout of images and text, and the size and other information.

After designing the original template, we will expand the original template library by means of color migration and size expansion. The richness and quantity of the template library are related to the deisgn effect. When the richness and quantity of the template library are higher, the synthesis effect will be better, and a better layout can be matched.

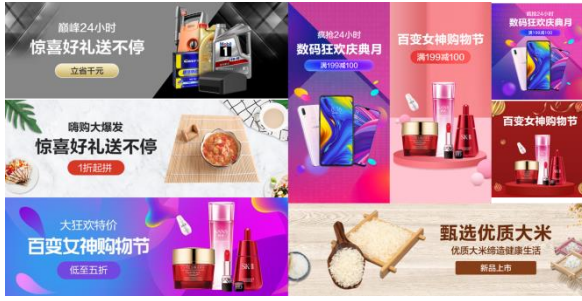 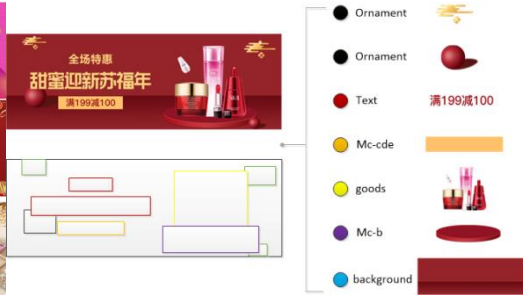

fig 2. template library  fig 3. layer type name design diagram

The templates in the template library shown in fig 1. Fig 2 shows the major categories of layer names. In fact, for the decorative layer, we have ornment-a-1 (trim layer) and ornment-b-1 (non trim layer). For mask layers, matching with different text layers produces different mask types. For example, matching with the main text is called mc-c, matching with sub-text is called mc-d, matching with action word is called mc-e, etc.

3.1.2 template recall

Template recall can have multiple ways. At present, our main recall method is to combine feature dimension screening and layout screening.

a. Feature dimension screening includes style, category and size information, and matches the closest size from the input information, and will further fine-tune the size later to match similar styles and sizes.

b. For the template of feature dimension recall, we calculate the input text and product information, and design the loss function of text information, and finally get a weighted loss, so as to recall the optimal template.

The loss function of the text information is

$$\text{score} = 1 - (\alpha_1 * \text{loss}_1 + \alpha_2 * \text{loss}_2)/\alpha_3$$

where loss1 is the difference between the number of lines of input text and the number of lines of template text, which may be 0, 1, 2. loss2 is the difference between the average word count of text in the template minus the average word count of the input text. among $\alpha_1$, $\alpha_2$ and $\alpha_3$ the three parameters can be derived from theory. When we restrict text, we stipulate that the minimum number of words(wordnummin) is 4 and the maximum number of words(wordnummax) is 20. When the number of lines is the same, we hope to give priority to the template with a little more words on average. When loss2 is greater than 0, $\alpha_2$ is 2, and when loss2 is less than 0, $\alpha_2$=-2.2. When the number of input text lines is 1, the average number of words is y, there are 1, 2, and 3 lines of text in the template, and the average number of words is x1, x2, x3, and the loss of each template is:

$$\text{loss}_1 = \alpha_1 * 0 + \alpha_2 * (x_1 - y)$$
$$\text{loss}_2 = \alpha_1 * 1 + \alpha_2 * (x_2 - y)$$
$$\text{loss}_3 = \alpha_1 * 2 + \alpha_2 * (x_3 - y)$$

We want loss1<loss2<loss3, then $\alpha_1 > \alpha_2 * (x_1 - y) - \alpha_2 * (x_2 - y)$, then $\alpha_1 > \alpha_2*(x_1-y)$. Using the prior maximum number of words to be 20 and the minimum number of words to be 4, $\alpha\alpha_1 = 2.2 * (\text{wordnummax} - \text{wordnummax}) + 1$. $\alpha_3$ is used for normalization, which can be obtained by the formula: $\alpha_1 * (\text{textline} - 1) + 2.2 * (\text{wordnummax} - \text{wordnummin})$.

The product score is obtained by the matching relationship between the size of the product and the size

of the product in the template. The final total recall score is:

$$score = text\_score * text\_weights + goods\_score * goods\_weights$$

where text_weights and goods_weights can be obtained by experience.

3.1.3 template pre-ranking

The template pre-ranking mainly filters the color modules of the template, and the template color distribution is an important part of the template tonality. Usually, we will have multiple color pre-ranking strategies. Including pre-ranking by calculating the distance between the input product color and the product color in the original template, which depends on the color matching logic of the product and material designed by the designer. However, this approach is not completely reliable, because the style / main color of some templates is designed according to the text, and there is semantic information in the text. Therefore, we usually consider both the top 1-3 of the input product color and the weighting of the product layer color in the template. Of course, only through thepre-rankingof color dimensions is still relatively single, which can be sorted and considered from more dimensions.
In addition, in the color pre-ranking module, we also need to consider:
a. Enter the color clash analysis of the product and template.
b. Sort topk of recall templates, control the proportion of color migration from uniform templates, and ensure the diversity of recall templates.
c. Recalls the sorting topk of the template, avoiding the proportion of templates from the same color family.

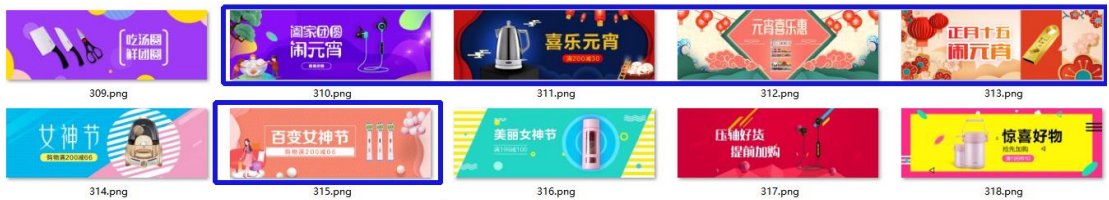

Fig4. the collocation of products and text styles in the pre-ranking

3.1.4 creative ranking

Compared with using artificial experience for creative selection, the inherent advantage of machine intelligence is that it can learn the advantages and disadvantages of creativity through massive data and accumulate knowledge online, so as to select the advertising creativity that can attract the attention of consumers most. Creative ranking is the optimal refinement of recall templates. Ranking is a supervised learning model, and the training data comes from historical click data. From the historical click data, by controlling the exposure and click thresholds(Usually, the exposure average and the click average in the bucketing period are taken. Those higher than the average value are used as positive samples, and those lower than the average value are used as negative samples.), we perform bucket processing on the CTR of the same template in different periods. Embedding of feature dimension includes image features, text features, recall/pre-ranking score, color and other dimensions.
a.image features. We trained the autoencoder self-supervised model, and also trained the supervised model that uses the divisions in the historically posted image data as label. The clustering effect of the autoencoder self-supervised model is very good, and finally we use the self-supervised training model to extract features.

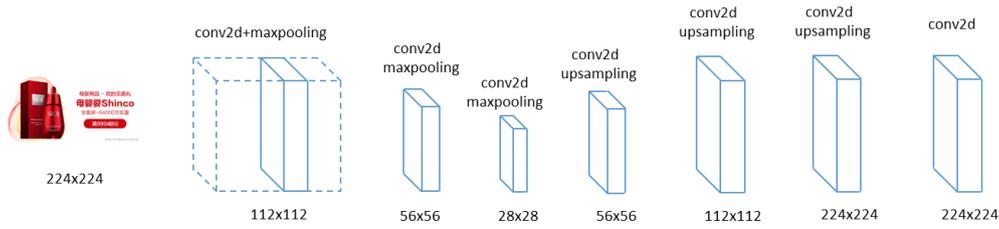

figure a

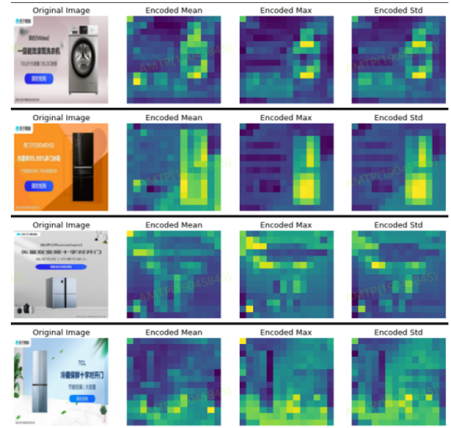

| layer | Output shape | param |
|---|---|---|
| conv2d_1 | None,224,224,32 | 896 |
| maxpolling2d_1 | None,112,112,32 | 0 |
| conv2d_2 | None,112,112,16 | 4624 |
| maxpooling2d_2 | None,56,56,16 | 0 |
| conv2d_3 | None,56,56,8 | 1160 |
| maxpooling_3 | None,28,28,8 | 0 |
| conv2d_4 | None,28,28,8 | 584 |
| upsampling2d_1 | None,56,56,8 | 0 |
| conv2d_5 | None,56,56,16 | 1168 |
| upsampling2d_2 | None,112,112,16 | 0 |
| Conv2d_6 | None,224,224,32 | 4640 |
| Upsampling2d_3 | None,224,224,32 | 0 |
| Conv2d_7 | None,224,224,3 | 867 |
| Total |  | 13939 |

figure b

figure c

Fig5 figa/b represents the network structure and parameters of the autoencoder, figc represents the feature cam map extracted by the autoencoder, and ae extracts a good feature dimension.

b.text. The text dimension uses tfidf as word frequency feature embedding. We also tried finetune bert[20] for downstream text feature extraction, but since the pattern of the text is relatively fixed and there is no long continuous semantic information, the bert effect of finetune is general. In the end, we only chose tfidf to extract word frequency. In addition, we did one-hot encoding for some words such as "discount" and "full discount" in the text.

c. The color is extracted from the main color, and 12 main colors are artificially designed, including black, white, red, orange, yellow, olivine, green, turquoise, grass, indigo, blue, purple, magenta,wine.

d. We considered numerical feature dimensions such as size and id. For example, seriesid is the template group number, id is the template number, and department is the business department number of the product group.

e. We also take the scores of template recall and template pre-ranking as part of the embedding dimension.

In the decision model, we also considered some multi-modal methods of graphics and text, but in the end we chose lgb[21], which is enough for creative optimization in our scene.

| Model | Feature emdedding | Model | Accuracy | AUC |
|---|---|---|---|---|
| lgb1 | seriesid,id,text,color,size,score | lgb | 0.6524 | 0.705459 |
| lgb2 | seriesid,id,text,tfidf,color,size,score | lgb | 0.6569 | 0.714433 |
| lgb_ae | seriesid,id,tfidf,ae,size | lgb | 0.6829 | 0.737812 |
| lgb3 | seriesid,id,text,tfidf,size,score,color,dullness | lgb | 0.6444 | 0.69856 |
| lgb_arcface | arcface,tfidf | lgb | 0.6041 | 0.630242 |
| lgb4 | seriesid,id,text,tfidf,color,size,score,department | lgb | 0.6621 | 0.721802 |
| lgb5 | seriesid,id,seriesid_id,text,tfidf,color,size,score，department | lgb | 0.6608 | 0.718033 |

Tab 1. Comparison of acc and auc under different dimensions of ranking

### 3.1.5 datasets

In the In the smartbanner framework, there are multiple sets of banner datasets, which are concentrated in the planner. In the template recall, a template library is designed, and the design of the template library can be extended to specific business scenarios. At present, we have several sets of template libraries, one is the banner template library, there are about 1200 banners, including different sizes and aspect ratios, etc., and the styles are relatively rich. There is also an atmosphere map template library, which mostly focuses on the atmosphere map of the big promotion scene and the social sharing map template library on the applet side. In the creative ranking, the data for our offline training currently comes from the data of the website's internal and external investment information flow. We screen according to the cycle, and the average training sample is about 70w.

### 3.2 actuator

The actuator mainly includes two aspects, including the size expansion module for obtaining the target size, the text layout and the product layout template. After the planner gets a good layout, we need to rearrange the input text area. After entering multiple products, we also need to combine product areas to meet the needs of advertisers.

### 3.2.1 size expansion

In the planner, we recall the size dimension, and the recalled templates are templates that are close to the target size. But it is not equal to the target size. For the target size, we need to expand the size of the template to obtain the target size. The size extension tool mainly relies on topological relationships for design. In fact, the size of the recalled template and the target size are relatively close, in this case, the size expansion can be very good(in a space with an aspect ratio of 30 degrees).

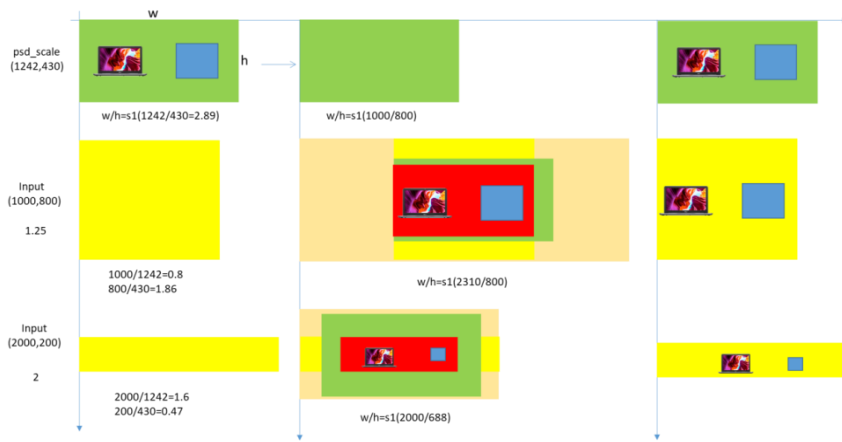

Fig6 Schematic diagram of topological relationship conversion for size expansion

### 3.2.2 text layout

Text module we design a supervised model. Based on the collection and analysis of existing template data, seven types of typesetting for all texts are specified. Normalize the input W, which is usually the length of the template. In addition, the input also includes character type encoding, such as character case or Chinese and English information encoding. The model outputs 2 heads, where the first head is responsible for predicting the w of the output text, and the second head is responsible for predicting the typesetting type of the output. Later, the layout of the text area will be reconstructed according to the output w and typesetting type.

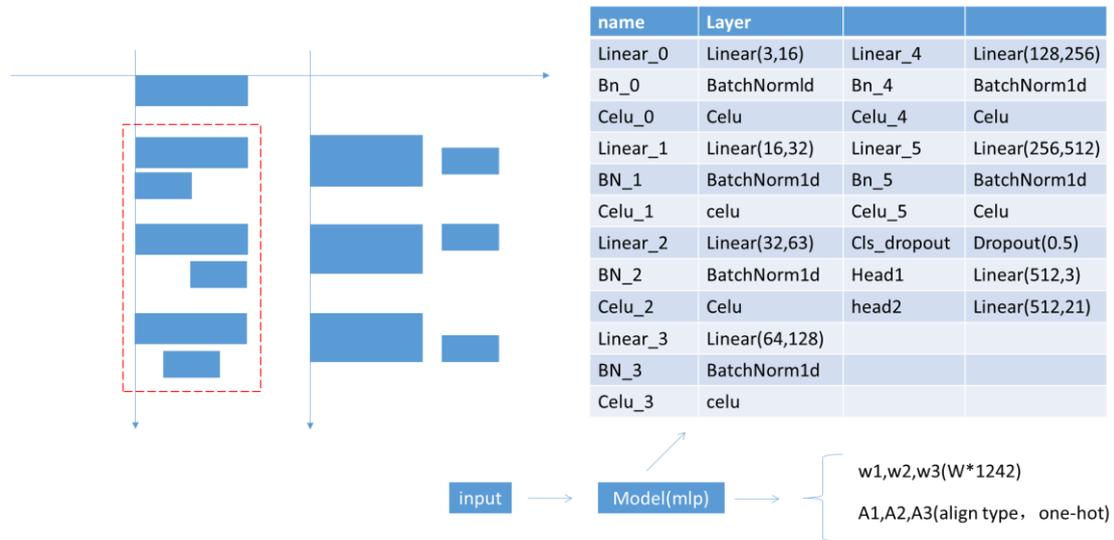

Fig7. The coordinates are the 7 forms of text, and the right side is the mlp model of text

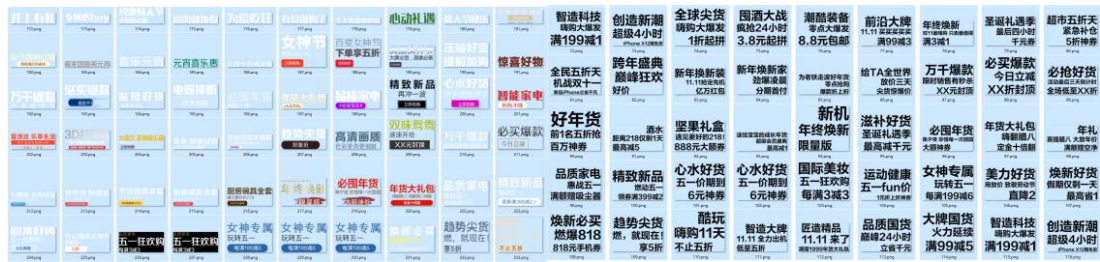

Fig8. The left figure is the layout of template text, and the right figure is the layout of model prediction output.

3.2.3 product layout

For the layout of multi product areas, we design a priori knowledge base according to the layout priori provided by the designer, and select the product layout according to the number of input products. When the number of input products is 2, there are three layouts: diagonal layout, inclined layout, side-by-side layout. When the number of input items is 3, there are 5 layouts, three-width layout, one-height two-width layout, three-slender layout, three-height layout, and triangular layout. When the input product is greater than 3, there are four layouts, the near-vision rotation layout, the two-layer triangle layout, the fan-shaped rotation layout, and the one-layer triangle layout. The specific form is to choose the appropriate layout method according to the aspect ratio of the product.

3.3 adjuster

The adjuster is the last step in the output template, similar to the goalkeeper role on a football field. In the dynamic fine-tuning link, we believe that the style of the design has been basically determined, referring to the designer's creation process, some adjustments need to be made to further meet the aesthetic standards.

The adjuster is structurally responsible for two parts of the work. The first part is semantic element retrieval. In order to better adapt to the changes of product and text area and size after input reconstruction, and ensure the coordination and unity of banner in semantic information. We analyze the layer information in the original template, learn from the designer's conventional design concepts and design rules, add product or text modification layers according to the banner situation, and build a template matching algorithm to select and replace the material elements consistent with the reconstructed banner semantic information from the picture template library.

The second part is to maintain a collection of fine-tuned algorithms. This optimization process is implemented by reinforcement learning, and action set includes movement, layer order adjustment, color change, scaling, transparency adjustment, etc. The evaluation of aesthetics and effects adopts some basic design principles such as whether the layout and color matching are reasonable and whether there is occlusion. At the same time, it also combines the evaluation criteria of some business characteristics such as concise background, prominent main content, and coordinated text lines and proportions. By controlling the design and configuration of the evaluation function, we are compatible with the business characteristics of different scenarios, so that the restricted configuration rules and the general aesthetic rules are naturally integrated. It can ensure the uniformity of colors, the visual impact of advertising images, and the neatness and unity of some complex alignment rules.

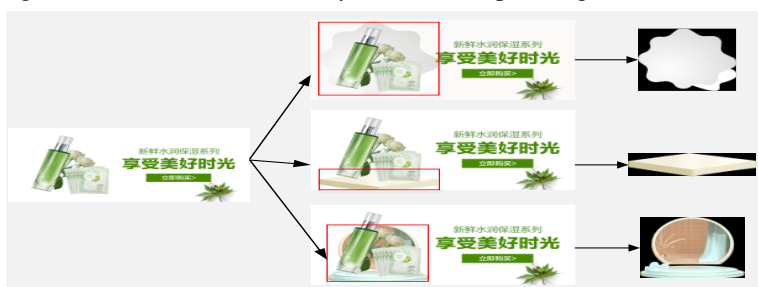

Fig9. Semantic element retrieval diagram, replacing the decoration layer of the product in it

3.4 evaluator

The evaluation of images is relatively subjective, and can be evaluated mainly from two aspects: aesthetics and effects. The aesthetic point of view can include lower-level judgment criteria such as alignment, reasonable color matching, and occlusion, as well as higher-level criteria, such as whether the style is consistent and whether it fits the theme. In terms of effect, whether the click-through rate will be improved after the product is launched. We have established a new module intelligent optimization, which is responsible for the generation of traceability templates and the regression of the final click rate. The generated nodes and recommendation feature embeddings are combined to form a recommendation model, thereby improving ctr.

4.Results

Smartbanner is used in multiple scenarios, and with the change and enrichment of the template library, various banner diagrams can be synthesized. Smartbanner has been practiced in many scenarios in Suning. Deployed a cluster of 48 cpu machines in the production environment, using Intel(R) Xeon(R) Silver 4114 CPU @ 2.20GHz 8G10C, centos7.3. The generator takes the most time in the smartbanner, especially when faced with a multi-layered banner, the compositing speed is positively related to the number of layers. In an average 7-layer banner, the samrtbanner single image synthesis takes about 200ms, and we parse the template psd in advance to form the smartbanner data format.We show a composite image of part of the smartbanner as shown below. And we show the results of the comparison between Luban and Linglong (we used the synthetic interface test of Luban and Linglong official website). Linglong only supports one-line and two-line text input and valid size, Luban can support one- to three-line multi-line text input, but also supports limited size, smartbanner can support one to three lines of input and unlimited output size. From the synthesis results, it can be seen that the color matching of Linglong and Luban are relatively incongruous, and smartbanner can optimize this through rough arrangement. Linglong and Luban's text layout is also relatively rigid, and smartbanner can better solve one point through the text layout model.

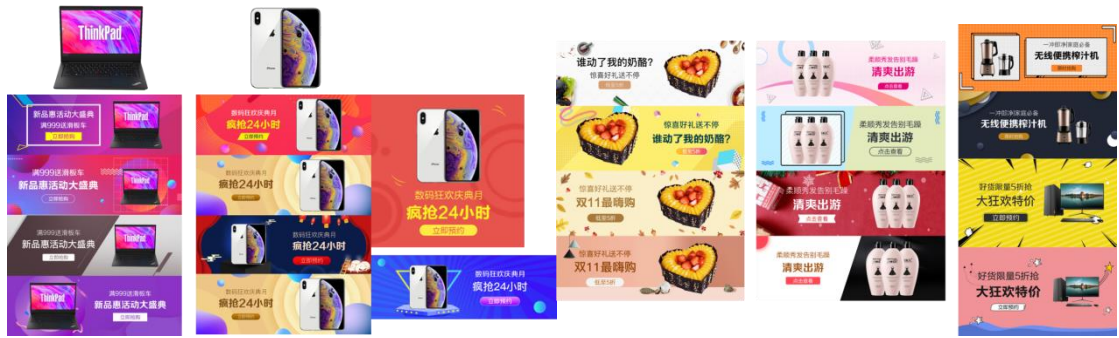

Fig10. smartbanner's smart design

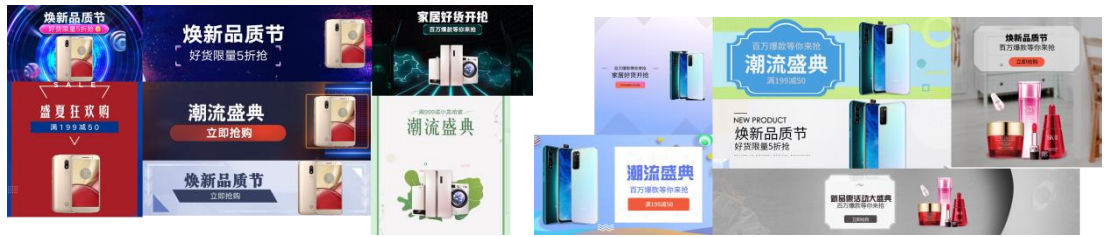

Fig11. On the left is the banner of Linglong, and on the right is the banner of Luban

5.conclusion

Smartbanner consists of planners, movers, adjuster and generators. Synthesize compliant banners with only a few creative elements in the input. The input of a small number of creative elements means that the smartbanner itself has a high degree of freedom. Modules such as layout and fine-tuning can relatively generate uncontrolled information, which increases the risk of banner unavailability. It is unacceptable in the scale image design of e-commerce. Smartbanner applies a lot of a priori knowledge, which can effectively reduce and control the degree of creative freedom, taking into account the degree of creative freedom and design rules, which not only ensures good creativity, but also ensures the standardization of design. Smartbanner fully liberates designers, lowers the threshold and cost of design, realizes rapid and rich visual design in batches, provides thousands of users and thousands of faces, and promotes the large-scale implementation of intelligent design in the field of e-commerce.